Precision Cosmology: Successes and Challenges
Joel R. Primack
*Physics Department, University of California, Santa Cruz, CA 95060 USA*

After briefly reviewing the good agreement between large-scale observations and the predictions of the now-standard ΛCDM theory and problems with the MOND alternative, I summarize several of the main areas of possible disagreement between theory and observation: galaxy centers, dark matter substructure, angular momentum, and the sequence of cosmogony, updating earlier reviews [1]. All of these issues are sufficiently complicated that it is not yet clear how serious they are, but there is at least some reason to think that the problems will be resolved through a deeper understanding of the complicated "gastrophysics" of star formation and feedback from supernovae and AGN.

## 1. Cold Dark Matter has come of age

It was once commonly assumed that the early universe was complex. But starting in the 1970s, Zel'dovich and his colleagues assumed that it was fundamentally simple, with just a scale-free spectrum of small adiabatic fluctuations of (a) baryons. And then (b) neutrinos (hot dark matter) when that failed, because the improving upper limit on fluctuations in the cosmic microwave background (CMB) ruled out a baryon-dominated universe by about 1980, and physicists in Moscow claimed to detect a mass for the electron neutrino of ~ 30 eV. My colleagues and I thought simplicity a good approach, but we tried other candidates for the dark matter, first supersymmetric (gravitino) warm dark matter [2], and then cold dark matter (CDM), hypothetical particles that moved sluggishly in the early universe. CDM led to a natural explanation for the observed masses and other properties of galaxies and clusters [3]. It soon became clear [4] that only two versions of CDM agreed with the rapidly improving data, ΛCDM – the "Double Dark" theory [5] – with about 25% of the density in cold dark matter and 70% in a some form of dark energy such as a cosmological constant (see Figure 1), and CHDM with about 75% of the density in cold and 20% in hot dark matter. The ΛCDM theory has been strongly favored since 1998, when observations of high-redshift Type Ia supernovae provided evidence that the expansion of the universe has been speeding up for about the past five billion years. Much more evidence is now available, especially from CMB [6] and large-scale structure observations [7], and the agreement with the predictions of ΛCDM is spectacular.

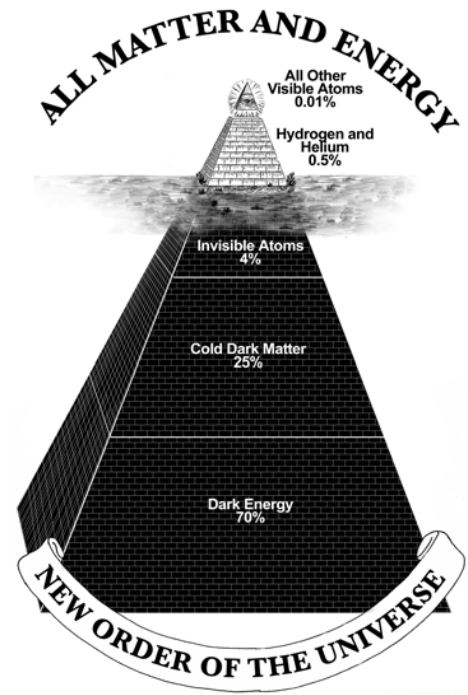

**Fig.1** [8]

## 2. Alternatives to CDM: MOND

Since the dark matter has not yet been detected and its nature remains a mystery, astrophysicists have naturally considered alternative explainations for the data. No such alternative has yet emerged that is remotely as successful as ΛCDM, but one that has attracted attention is modified Newtonian dynamics (MOND) [9]. Besides the fact that MOND is not a predictive cosmological theory and has problems explaining gravitational lensing observations even on galactic scales [10], I want to call attention to two sorts of data that strongly disfavor MOND and support CDM. One is the common observation of galaxies in late stages of merging, in which the dense galactic nuclei have nearly coalesced while the lower-density surrounding material is still found in extended tidal streamers. This is exactly what happens in computer simulations of galaxy mergers, in which the process of dynamical friction causes the massive nuclei to lose kinetic energy to the dark matter and quickly merge. But if there were no dark matter, there would be nothing to take up this kinetic energy and the nuclei would continue to oscillate for a long time. The second sort of data comes from studies of galaxy clusters [11], including their aspherical shapes which agree well with the predictions of ΛCDM [12]. The "bullet" cluster 1E0657-56 shows particularly clearly that the cluster baryons account for only a small part of the mass, contrary to MOND [13], and also disfavors [14] the "interacting dark matter" idea.

## 3. Challenges to CDM

The main challenges to the CDM paradigm are on small scales, and they include the following issues:

- *Cusps* – too much dark matter in halo centers?
- *Not enough dark matter* around elliptical galaxies?
- Halo *substructure* issues
- *Angular momentum* issues
- Halo and galaxy merging history and the *sequence of cosmogony*

The *cusp* problem – the possible disagreement between the ~1/r dark matter density profile predicted by simulations and dwarf galaxy rotation curve measurements – was first pointed out in 1994 [15]. But it was caused in part by theorists (including your author) not understanding that the error bars on the neutral hydrogen (HI) radio data were unrealistically small. When the large size of the radio beam was taken into account, much of the discrepancy was removed [16]. The latest high-precision data [17], using CO and Hα as well as HI, agree well with the expectations from ΛCDM when the effects of the nonspherical dark matter halos [12] are included [18]. The cusp problem now appears to be resolvable in CDM, although not everyone agrees that it has been solved. Meanwhile, the mean density $\Delta_{V/2}$ inside the radius $r_{V/2}$ where the rotation velocity reaches half the maximum observed value appears to be somewhat smaller than the ΛCDM prediction with dark matter linear density fluctuation power spectrum $P(k)$ normalization $\sigma_8 = 0.9$, but more consistent with $\sigma_8 = 0.75$ [19] favored in the latest WMAP paper [6].

The *not enough dark matter* problem arises from the observed rapid fall of the velocity dispersion of tracer stars around several otherwise typical elliptical galaxies, which was argued to imply that there is no dark matter around these galaxies [20]. If true, this would be very difficult to understand within the CDM paradigm. However, my colleagues and I found that these observations are actually in good agreement with the predictions of galaxy merger simulations in the usual ΛCDM framework, and others subsequently confirmed this both theoretically and observationally [21].

One of the main halo *substructure* issues concerns galactic satellites. The number of subhalos rises inversely as a function of subhalo mass, and greatly exceeds the number of observed satellite galaxies at masses below those of the Magellanic Cloud satellites of the Milky Way [22]. However, it is very plausible that the effects of photoionization and supernova feedback suppress the formation of visible galaxies in these small dark matter halos [23]. Indeed, theories such as warm dark matter which supress the formation of small halos are unlikely to form as many satellite galaxies as are observed, and such theories are also constrained by observations of the Lyα forest of small gas clouds and of galaxies and bright quasars at high redshifts [24].

By identifying the halos likely to host visible galaxies as those that had the largest masses or circular velocities before they fell into the more massive halos in which they are now satellites, it has been possible to explain key observed features of the observed satellite population including their radial distribution [25] and orientation [26]. The highest resolution simulations [27] show that the number of subhalos continues to rise with decreasing subhalo mass, inplying that dark matter annihilation within subhalos may be important both locally and in building up a cosmic background of gamma rays over cosmic time if the dark matter is supersymmetric neutralinos or other particles that can self-annihilate. Understanding anomalous ratios of image brightness in radio observations of gravitationally lensed quasars may require dark matter substructure, but the relevant constraints on halo substructure are not yet clear since some of the lensing may be caused by intervening objects [28].

The growth of the mass of dark matter halos and its relation to the structure of the halos has been studied based on structural merger trees [29], and the *angular momentum* of dark matter halos is now understood to arise largely from the orbital angular momentum of merging progenitor halos [30]. But it is now clear that the baryonic matter in disk galaxies has an angular momentum distribution very different from that of the dark matter [31]. Although until recently simulations were not able to account for the formation and structure of disk galaxies, simulations with higher resolution and improved treatment of stellar feedback from supernovae are starting to produce disk galaxies that resemble those that nature produces [32]. It remains to be understood how the gas that forms stars acquires the needed angular momentum. Possibly important is the recent realization that gas enters lower-mass halos cold and in clouds [33], rather than being heated to the halo virial temperature as in the standard treatment used in semi-analytic models [34].

The *sequence of cosmogony* is the order in which structures form in the universe – for example, galaxies and then stars. One of the reasons that many observers have remained suspicious of CDM is that it seems to predict that small galaxies form first and then larger ones afterward, while it seems observationally that the largest galaxies have the oldest stars, with smaller galaxies having younger stellar populations. This is one of the related phenomena that the observers call "downsizing" [35]. But downsizing is not necessarily contrary to CDM. It is true that the CDM power spectrum first falls slowly as a function of the mass enclosed in a spherical volume, and then more rapidly as the volume encloses more mass than is contained in a large galaxy cluster [3]. As a result, CDM is a hirerarchical cosmology, in which small dark matter halos form early, then are mostly engulfed within larger collapsing halos, and so on to still larger scales. But all the visible material in galaxies represents only about 0.5% of the total density of the universe (represented by the small above-ground part of the cosmic density pyramid in Fig. 1), and its dynamics reflects a complex interplay between the gravity of the dark matter and the energy generated and transferred through gas dynamics, nuclear processes in stars, and the infall of matter into black holes fueling active galactic nuclei (AGNs).

Mergers between disk galaxies have long been thought to be responsible for forming massive stellar spheroids – elliptical galaxies and the bulges of large diskl galaxies – which harbor a large majority of the stellar mass in the present universe. Until recently, however, simulations and semi-analytic models, including those with which I have been associated [36], did not predict the galaxy color bimodality – the division of galaxies into blue ones with active star formation and red ones without – which is now clearly seen both nearby and out to redshift ~1.4 [37]. But these models did not include the effects of hydrodynamic process [38] and AGN [39] which are likely to be important in suppressing star formation after galactic mergers. A concern is that ΛCDM predicts too much galaxy merging, particularly if $\sigma_8$ is as low as 0.75, which will destroy galactic disks and prevent the formation of the large population of disk galaxies seen today, especially the lower mass disk galaxies with small or absent bulges. This potential problem may be alleviated by the lower rate of merging of lower-mass galaxies in the latest cosmological hydrodynamic simulations [40].

Recently, a major program of galaxy merger simulations has been run at Harvard including the effects of AGN, with the constraint that the resulting black hole mass is about 1/1000 that of its host stellar spheroid, consistent with observations of nearby galaxies [41]. This has led to a remarkable theory [42] that connects such mergers to the evolution of the quasar luminosity function and the formation of ultra-luminous infrared galaxies (ULIRGS) which evolve into bright elliptical galaxies. Meanwhile, my colleagues and I are continuing our own large program of galaxy merger simulations including the effects of dust, which typically absorbs ~90% of the light and reradiates it at far-infrared wavelengths [43]. We have developed new techniques for identifying mergers using galaxy optical morphology [44], and we are now comparing the resulting data including AGN [45] to simulations. Work is progressing in understanding the properties of the galaxies produced by such mergers [46] and comparing them systematically [47] with observations such as those of galaxy kinematics using the SAURON integral field spectrograph [48]. Semi-analytic models incorporating rapid quenching and continued supression of star formation after mergers are now succeeding in accounting for the observed galaxy color bimodality and many other observed properties of the evolving galaxy population [49].

## 4. Summary and Prospect

- We now know the cosmic recipe. Most of the universe is invisible stuff called "nonbaryonic dark matter" (~25%) and "dark energy" (~70%). Everything that we can see makes up only about 0.5% of the cosmic density, and invisible atoms about 4%. The earth and its inhabitants are made of the rarest stuff of all: heavy elements (0.01%).
- The ΛCDM "Double Dark" theory based on this appears to be able to account for all the large scale features of the observable universe, including the details of the heat radiation of the Big Bang and the large scale distribution of galaxies.
- Constantly improving data are repeatedly testing this theory. The main ingredients have been checked several different ways. There exist no convincing disagreements, as far as I can see. Possible problems on subgalactic scales may be due to the poorly understood physics of gas, stars, and massive black holes.
- But we still don't know what the dark matter and dark energy are, nor really understand how galaxies form and evolve. There's lots more work for us to do, much of was discussed at this meeting. With the constantly improving sensitivity of direct and indirect dark matter searches, the possibility that supersymmetric dark matter particles might be produced as early as next year at the Large Hadron Collider, and the careful study of alternative methods of discovering the nature of the dark energy by the Dark Energy Task Force, our prospects for progress are very bright!

## Acknowledgments
I acknowledge support from NASA and NSF grants, and also from STScI Theory grants. I also thank my students and other collaborators for many helpful discussions, and David Cline for inviting me to give this talk.